\documentclass[twocolumn,aps,prb,showpacs]{revtex4}

\usepackage{epsfig}

\newcommand{\beq}{\begin{equation}}
\newcommand{\eeq}{\end{equation}}
\newcommand{\bea}{\begin{eqnarray}}
\newcommand{\eea}{\end{eqnarray}}

\begin{document}

\title{Signatures of non-monotonic $d-$wave gap in electron-doped cuprates}
\author{Ilya Eremin$^{1,2}$, Evelina Tsoncheva$^{3}$,
 and  Andrey V. Chubukov$^3$}

\affiliation{$^1$ Max-Planck Institut f\"ur Physik komplexer
Systeme, D-01187 Dresden, Germany \\
$^2$ Institute f\"ur Mathematische und Theoretische  Physik,
TU-Braunschweig,
D-38106 Braunschweig, Germany \\
$^3$ Department of Physics, University of Wisconsin,
Madison, WI 53706}
\date{\today}

\begin{abstract}
We address the issue whether the data on optical conductivity and
Raman scattering in electron-doped cuprates below $T_c$ support the
idea that the $d-$wave gap in these materials is non-monotonic along
the Fermi surface. We calculate the conductivity and Raman intensity
for elastic scattering, and find that a non-monotonic gap gives rise
to several specific features in optical and Raman response
functions. We argue that all these features are present in the
experimental data on Nd$_{2-x}$Ce$_{x}$CuO$_4$ and
Pr$_{2-x}$Ce$_{x}$CuO$_4$ compounds.
\end{abstract}

 \maketitle

\section{introduction}

The studies of electron-doped cuprates,
Nd$_{2-x}$Ce$_x$CuO$_{4-\delta}$ (NCCO)
 and Pr$_{2-x}$Ce$_x$CuO$_{4-\delta}$ (PCCO) are
attracting considerable attention from high-$T_c$ community. The
phase diagram of  electron-doped cuprates is not as involved as in
hole-doped materials. It contains sizable regions of
antiferromagnetic and superconducting phases, and only a small
region showing pseudogap behavior~\cite{millis}. The superconducting
dome is centered around a quantum-critical point at which the
antiferromagnetic $T_N$ vanishes, in close similarity to phase
diagrams of several heavy-fermion materials~\cite{rev_hf}.

Scanning SQUID~\cite{tsuei} and ARPES
experiments~\cite{matsui,armitage} on electron-doped cuprates
provided strong evidence that the  gap symmetry is $d_{x^2-y^2}$,
same as in hole-doped cuprates. This gap has nodes along the
diagonals of the Brillouin zone, and  changes sign twice along the
Fermi surface. The functional form of the $d_{x^2-y^2}$ gap is a
more subtle issue, however. In hole-doped cuprates, the  gap
measured by ARPES follows reasonably well a simple $d-$wave form
$\Delta (k) = \frac{\Delta_0}{2} (\cos k_x - \cos k_y)$ (equivalent
to $\cos 2 \phi$ for a circular Fermi surface), at least near and
above optimal doping~\cite{icc}. In the electron-doped cuprates,
high-resolution ARPES data on the leading-edge gap in
Pr$_{0.89}$LaCe$_{0.11}$CuO$_4$ (Ref.~\cite{matsui}) show a
non-monotonic gap, with a maximum in between nodal and antinodal
points on the Fermi surface.  Such gap was earlier proposed in
Ref.~\cite{blumberg} as a way to explain Raman experiments in NCCO,
particularly the higher frequency  of the pair-breaking, '$2\Delta$' peak
in the $B_{2g}$ channel than in the $B_{1g}$
channel. Recent measurements of optical conductivity $\sigma_1
(\omega)$ in Pr$_{1.85}$Ce$_{0.15}$CuO$_4$ (Ref. \cite{homes}) were
also interpreted as an indirect evidence of a non-monotonic gap.

The interpretation of the experimental results is still
controversial, though. ARPES data on PCCO below $T_c$ in
Ref.\cite{matsui} show a non-monotonic leading-edge gap, but the
spectral function all along the Fermi surface does not display a
quasiparticle peak, from which one would generally infer the
functional form of the gap more accurately. The interpretation of
the Raman data has been criticized in Ref. \cite{rudi} on the basis
that, within BCS theory, the shapes of $B_{1g}$ and $B_{2g}$ Raman
intensities for the non-monotonic gap proposed in Ref.~\onlinecite{blumberg}
do not agree with the data.  Finally, optical results for PCCO in
Ref.~\onlinecite{homes} do show a maximum at about $70 meV$, which is close
to $2\Delta_{max}$ inferred from $B_{2g}$ Raman
scattering. However,  it is {\it a-priori} unclear whether one should
actually expect such maximum in the optical conductivity. In
particular, in hole-doped materials, $\sigma_1 (\omega)$ is rather
smooth at $2\Delta_{max}$~\cite{basov_rev}.

\begin{figure}[h]
\epsfig{file=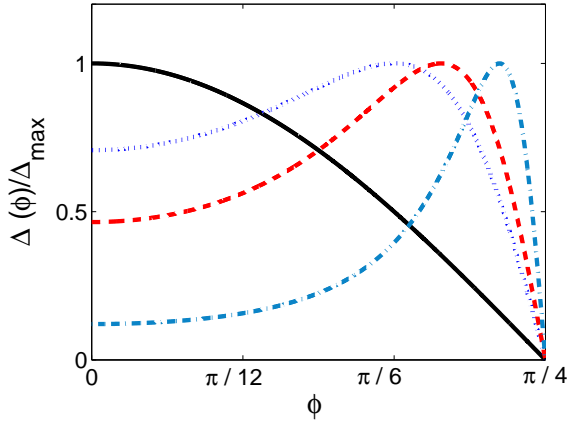,width=8.cm} \caption{(color online) The
non-monotonic $d_{x^2-y^2}$-wave gap calculated using
Eq.(\protect\ref{r_2}) for $a=2$ (dotted curve), $a=4$ (dashed
curve), and $a=20$ (dashed-dotted curve). The black straight curve
is a monotonic $d_{x^2 - y^2}$-wave gap, $\Delta_\phi =
\Delta_{max} \cos 2 \phi$.} \label{fig2}
\end{figure}

From theory perspective, the non-monotonic $d-$wave gap appears
naturally under the assumption that the $d_{x^2-y^2}$ pairing is
caused by the interaction with the continuum of overdamped
antiferromagnetic spin fluctuations. Spin-mediated interaction is attractive
in the $d_{x^2-y^2}$ channel and yields a gap which is maximal near
the hot spots -- the points along the Fermi surface, separated by
antiferromagnetic momentum, {\bf Q}$_{AF}$. In optimally doped NCCO
and PCCO, hot spots are located close to Brillouin zone diagonals,
and one should generally expect the $d_{x^2-y^2}$ gap to be
non-monotonic \cite{manskerembenn}. More specifically, in the spin
fluctuation scenario, the maximum of the gap is slightly shifted
away from a hot spot towards antinodal region, such that $d-$wave
superconductivity with a non-monotonic $d_{x^2-y^2}$ gap survives
even when the hot spots merge at the zone diagonals~\cite{krot}. The
solution of the gap equation  in this case yields
\beq \Delta_\phi = \Delta_{max}~
 \left(\frac{2\sqrt{a}}{3\sqrt{3}}\right) \frac{\cos{2\phi}}{
(1 + a \cos^2{2\phi})^{3/2}}.
\label{r_2}
\eeq
Here, $\phi$ is the angle along the (circular) Fermi surface ($\phi
=\pi/4$ corresponds to a diagonal Fermi point), $a >1/2$ is a
model-dependent parameter, and $\Delta_{max}$ is the maximum value of the gap
 located at $\cos 2\phi = (1/2a)^{1/2}$.  The
gap at various $a$ is  shown in Fig.\ref{fig2}.
As $a$ increases, the nodal velocity increases, the maximum of the
gap shifts towards the zone diagonal, and the value of the gap at
the antinodal point $\phi =0$ decreases. A similar functional form
of the gap can be obtained by adding higher harmonics
$\cos{6\phi},~\cos{10\phi}$ etc. to the $\cos 2\phi$ gap. We have
found, however,  that Eq. (\ref{r_2}) is somewhat better for
experimental comparisons  than the gap with a few higher harmonics.
ARPES measurements~\cite{matsui} place the maximum of the gap
slightly below $\phi = \pi/6$. This is best
reproduced if we set $a=2$. However, since ARPES results have not
been yet confirmed by other groups, we will keep $a$ as a parameter
and present the results for various values of $a$.

The goal of our work is to verify to which extent optical
conductivity $\sigma_1 (\omega)$ and Raman scattering $R(\omega)$ in
a $d-$wave superconductor with a gap given by Eq. (\ref{r_2}) are
consistent with the experimental data.  For this, we computed
$\sigma_1 (\omega)$ and $R(\omega)$ in $B_{1g}$ and $B_{2g}$
geometries assuming that the scattering is elastic. The latter does
not necessarily have to come from impurities -- scattering by
collective excitations in spin or charge channels is also dominated
by processes with small frequency transfers. For simplicity we
assume that the normal state damping rate is independent on
frequency and only focus on the effects associated with the pairing.

We have found several features which distinguish optical and Raman
responses in superconductors with non-monotonic $d$-wave gap from
superconductors with a $\cos 2\phi$ gap. Optical conductivity in a
pure $d-$wave superconductor with elastic scattering has a weak
maximum followed by a broad suppression region at frequencies of the
order $\Delta_{max}$ (Ref. \cite{basov_rev}). For the non-monotonic
gap, we have found a rather strong maximum in $\sigma_1 (\omega)$
slightly below $2\Delta_{max}$, followed by a sharp drop in
conductivity down to very low frequencies, where the conductivity
begins to increase again towards a constant value at $\omega = 0+$
(see Fig. \ref{fig5_1}). For Raman scattering, we have observed that
the peak in the $B_{2g}$ channel is located at a higher frequency
than in the $B_{1g}$ channel, and also that the shapes of the two
Raman profiles are very different -- the $B_{2g}$ peak is
near-symmetric, the $B_{1g}$ peak is very asymmetric with
shoulder-like behavior above the peak frequency. We argue that these
features are consistent with the experimental conductivity and Raman
data. From this perspective, our findings give additional support to
the idea that the $d_{x^2-y^2}$ gap in electron-doped cuprates is
highly non-monotonic.

We present the formalism in Sec. \ref{sec:form}, and the results in
Sec. \ref{sec:res}. In the latter section, we  also consider the
comparison with the data in more detail. The last section is the
conclusion.

\section{The formalism}
\label{sec:form}

We adopt a conventional strategy of analyzing  optical and
 Raman responses in non-$s-$wave superconductors with impurity
scattering~\cite{lee,scalapino,abr}. We assume that the scattering
originates from the $s-$wave component of the effective  interaction
(which includes the impurity potential), and gives rise to a
$k-$independent fermionic self-energy $\Sigma (\omega)$. The pairing
comes from a different,  $d-$wave component of the interaction. As
in earlier works\cite{lee,scalapino,abr}, we assume that the
$d-$wave anomalous vertex is frequency independent, and to a
reasonable accuracy can be replaced by  $\Delta (\phi)$ from Eq.
(\ref{r_2}). The time-ordered normal and anomalous fermionic Green's
functions in this approximation are given by
\bea G_{\bf k} (\omega,\phi) &=& \frac{{\tilde \omega} +
\epsilon_{\bf k}}{{\tilde \omega}^2 - \epsilon^2_{\bf k} -
\Delta^2_\phi}
\label{r_3_1} \quad, \\
F_{\bf k} (\omega,\phi) &=& \frac{\Delta_\phi}{{\tilde \omega}^2 -
\epsilon^2_{\bf k} - \Delta^2_\phi} \quad,  \label{r_3} \eea
where ${\tilde \omega} = \omega + \Sigma (\omega)$. The self-energy
is by itself expressed via the (local) Green's function via
\beq \Sigma (\omega) =  i \frac{\gamma G^L(\omega)}{C^2 + (G^L
(\omega))^2} \quad, \label{r_4} \eeq
where the local Green's function is
\beq G^L (\omega) = i \frac{4}{\pi^2} \int_0^{\pi/4} \int d
\epsilon_{\bf k} G_{\bf k} (\omega, \phi) \quad, \label{r_5} \eeq
( $G_L =
1$ in the normal state),
and the parameter $C$ is interpolated between $C >>1$ in the Born
limit, and $C <<1$ in the unitary limit.

Optical conductivity $\sigma_1 (\omega)$ and  Raman intensity
$R(\omega)$ are both given by the combinations of bubbles made out
of normal ($GG$) and anomalous ($FF$) Green's functions. Optical
conductivity is proportional to the current-current correlator,
while Raman intensity is proportional to the density-density
correlator weighted with angle-dependent Raman vertex factors
\beq
\gamma_{B_{1g}} \propto \cos 2\phi, ~~~\gamma_{B_{2g}} \propto \sin
2 \phi. \label{r_5_1}
\eeq
To a first approximation, $B_{1g}$ Raman scattering then gives
information about electronic states in the antinodal regions, near
$\phi =0$,  while  $B_{2g}$ scattering gives information about nodal
regions, near $\phi = \pi/4$.

The overall sign of the $FF$ contribution is different for $\sigma
(\omega)$ and $R(\omega)$,  as the running momenta in the side
vertices in the $FF$ term are ${\bf k}$ and $-{\bf k}$, between
which the current operator changes sign, but the density operator
remains intact. For a constant density of states, which we assume to
hold, the integration over $\epsilon_{\bf k}$ in $GG$ and $FF$
bubbles can be performed exactly, and yields
\begin{widetext}
\bea \sigma_1 (\Omega) = - \frac{\omega^2_{pl}}{4\pi}~ \frac{2}{\pi
\Omega}~\mbox{Im}\left[ \int_0^{\pi/4} d \phi \int d\omega
\frac{\sqrt{{\tilde \omega}^2_+ -\Delta^2_\phi} \sqrt{{\tilde
\omega}^2_- -\Delta^2_\phi} - {\tilde \omega}_+  {\tilde \omega}_- -
\Delta^2_\phi}{\sqrt{{\tilde \omega}^2_+ -\Delta^2_\phi}
\sqrt{{\tilde \omega}^2_- -\Delta^2_\phi} \left(\sqrt{{\tilde
\omega}^2_+ -\Delta^2_\phi}  + \sqrt{{\tilde \omega}^2_-
-\Delta^2_\phi}\right)} \right] \quad,
\label{r_6} \\
R_i (\Omega) = - \frac{\sqrt{4}}{\pi^2} R_0 \int_0^{\pi/4} d \phi
\gamma^2_i \mbox{Im} \left[ \int d\omega \frac{\sqrt{{\tilde
\omega}^2_+ -\Delta^2_\phi}  \sqrt{{\tilde \omega}^2_-
-\Delta^2_\phi} - {\tilde \omega}_+  {\tilde \omega}_- +
\Delta^2_\phi}{\sqrt{{\tilde \omega}^2_+ -\Delta^2_\phi}
\sqrt{{\tilde \omega}^2_- -\Delta^2_\phi} \left(\sqrt{{\tilde
\omega}^2_+ -\Delta^2_\phi}  + \sqrt{{\tilde \omega}^2_-
-\Delta^2_\phi}\right)} \right] \label{r_7} \quad. \eea
\end{widetext}
Here, $\sigma_1$ is the real part of the conductivity, $\omega_{pl}$
is the plasma frequency, the index $i$ labels the various scattering
geometries, ${\tilde \omega} = \omega + \Sigma (\omega)$,
$\omega_\pm = \omega \pm \frac{\Omega}{2}$, and $R_0$ is the
normalization factor for the Raman intensity.  The conductivity in a
superconductor also contains a $\delta (\Omega)$  contribution (not
shown) related to the superconducting order parameter.

In a ideal BCS superconductor with ${\tilde \omega} = \omega$, the
conductivity $\sigma_1 (\omega)$ vanishes, while Raman intensity is
given by~\cite{klein,devereaux}
\beq R_i(\omega)= R_0
\mbox{Re} \left< \frac{\gamma^2_i (\phi)
\Delta^2 (\phi)}{\omega \sqrt{\omega^2 - 4 \Delta^2
(\phi)}}\right>_{FS} \label{r_1}
\eeq
where  $<...>$ denotes the averaging over the Fermi surface. For a
pure $d-$wave gap, $B_{1g}$ Raman intensity scales as $\omega^3$ at
small frequencies~\cite{devereaux}, and diverges logarithmically at
$2 \Delta_{max}$. $B_{2g}$ intensity scales as $\omega$ at small
frequencies and has a broad maximum at around $1.6 \Delta_{max}$
(see Fig. \ref{fig5} and \ref{fig6} below).

\begin{figure}[h]
\epsfig{file=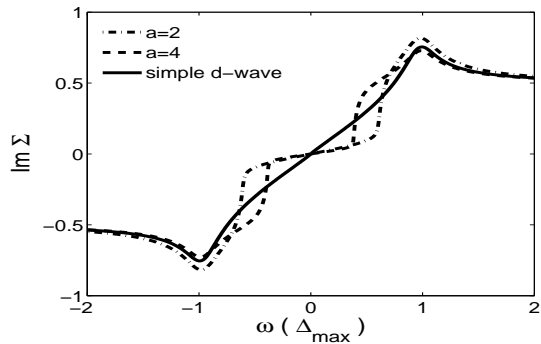,width=8cm,height=5cm} \caption{The behavior
of $Im {\tilde \omega} = Im (\omega + \Sigma (\omega) = Im \Sigma
(\omega)$ in the Born limit for a d-wave superconductor with a
monotonic gap (solid curve) and non-monotonic gap from Eq.
(\protect\ref{r_2}) with $a=2$ (dashed-dotted curve) and $a=4$
(dashed curve). We used ${\tilde \gamma} = \gamma/C^2 =0.05
\Delta$.}
 \label{fig3}
\end{figure}
\begin{figure}[h]
\epsfig{file=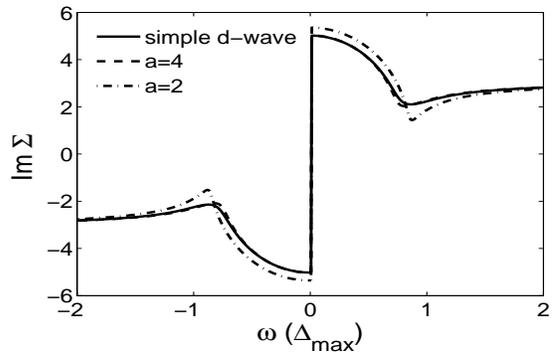,width=8cm,height=5cm} \caption{The
behavior of Im$\Sigma (\omega)$ in the unitary limit.  Solid curve
-- a monotonic $d-$wave gap, dashed and dashed-dotted curves -- a
non-monotonic gap from Eq. (\protect\ref{r_2}) with $a=4$ and $a=2$,
respectively. We used $\gamma =0.3 \Delta$.} \label{fig4}
\end{figure}

\section{The results}
\label{sec:res}

\subsection{Fermionic self-energy}

We computed the fermionic self-energy  by solving numerically the
self-consistent equation (\ref{r_4}) in the Born and unitary limits.
The results for the imaginary part of the self-energy in the Born
limit are presented in Fig.\ref{fig3}. For a $d-$wave superconductor
with a monotonic gap, Im$\Sigma$ is linear in frequency at small
$\omega$ and has a cusp at $\omega = \Delta_{max}$. For a
non-monotonic gap, Im$ \Sigma$ is reduced at small frequencies, and
then rapidly increases to a value comparable to that for a monotonic
gap. This behavior resembles, particularly for $a=2$, the formal
solution of Eq. (\ref{r_4}) for an angle-independent gap
$\Delta=\Delta_{max}$. In the latter case, Im$\Sigma =0$ up to a
frequency $\omega = \Delta (1 - ({\tilde
\gamma}/\Delta)^{2/3})^{3/2}$, where ${\tilde \gamma} = \gamma/C^2$,
and rapidly increases above this frequency.
For  ${\tilde \gamma} = 0.05 \Delta_{max}$, used in Fig. \ref{fig3},
the jump occurs at $\omega \approx 0.8 \Delta$, much like
 in the plot for $a=2$. We emphasize that the solution of (\ref{r_4})
 for a constant $\Delta$ is not the result for an $s$-wave
superconductor. For the latter, the fermionic self-energy and the
pairing vertex are renormalized by the same interaction, and the
self-consistent equation for $\Sigma (\omega)$ does not have the
form of Eq. (\ref{r_4}) with frequency independent $\Delta$.

In Fig. \ref{fig4} we show Im$\Sigma$ in the unitary limit $C=0$. We
observe the same trend. For a monotonic $d$-wave gap, Im$\Sigma$ is
nearly monotonic, and  has only a slight minimum around $0.8
\Delta_{max}$. For a non-monotonic gap, particularly for $a=2$, Im$
\Sigma$ has a more pronounced structure with a sharp minimum around
$0.8 \Delta_{max}$.  This behavior again  resembles that for a
constant gap $\Delta$. In the latter case, a formal solution of
(\ref{r_4}) for  $\gamma \ll \Delta$ yields a  zero Im$ \Sigma
(\omega)$ between $2 \sqrt{\gamma \Delta}$ and $\sqrt{\Delta^2 +
\gamma^2}$. At larger $\omega$, Im$\Sigma$ gradually approaches the
normal state value $\gamma$, at small frequencies it is also finite
and approaches $\sqrt{\gamma \Delta}$ at zero frequency [for a
generic $C$, a non-zero $Im \Sigma (\omega =0)$ (the unitary
resonance) appears when $\gamma$ exceeds $\Delta C^2 \sqrt{1+
C^2}$]. The region of vanishing Im$\Sigma$ shrinks to zero when
$\gamma$ exceeds the critical value of $2\Delta/(3\sqrt{3}) \approx
0.4 \Delta$. For  the same $\gamma =0.3 \Delta_{max}$ as used in
Fig. \ref{fig4},  Im$ \Sigma$ for a constant gap sharply drops
around $1.1 \Delta_{max}$, and  rebounds both at larger and smaller
frequencies, much like our actual solution for $a=2$.

\begin{figure}[h]
\epsfig{file=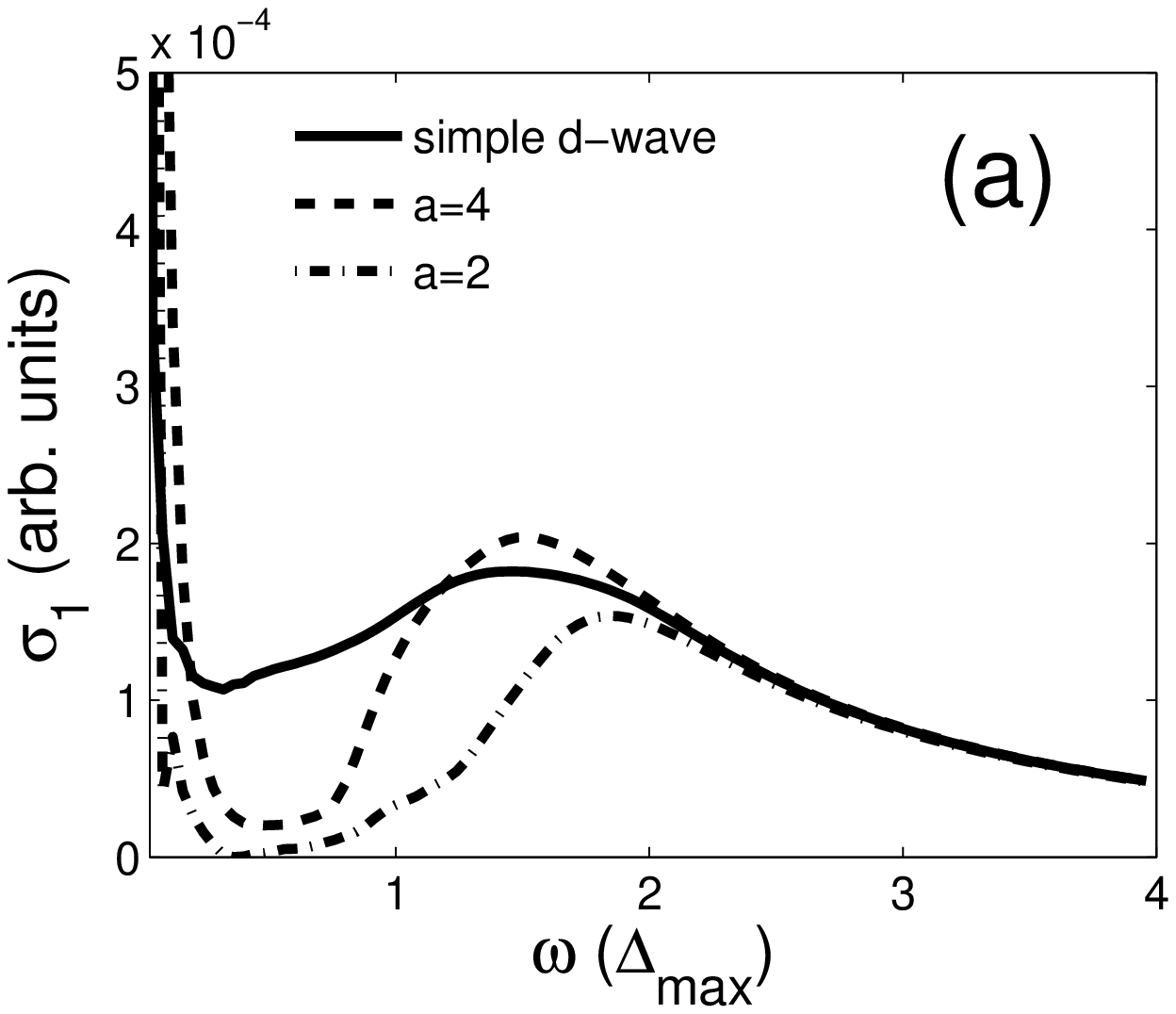,width=6cm}
\epsfig{file=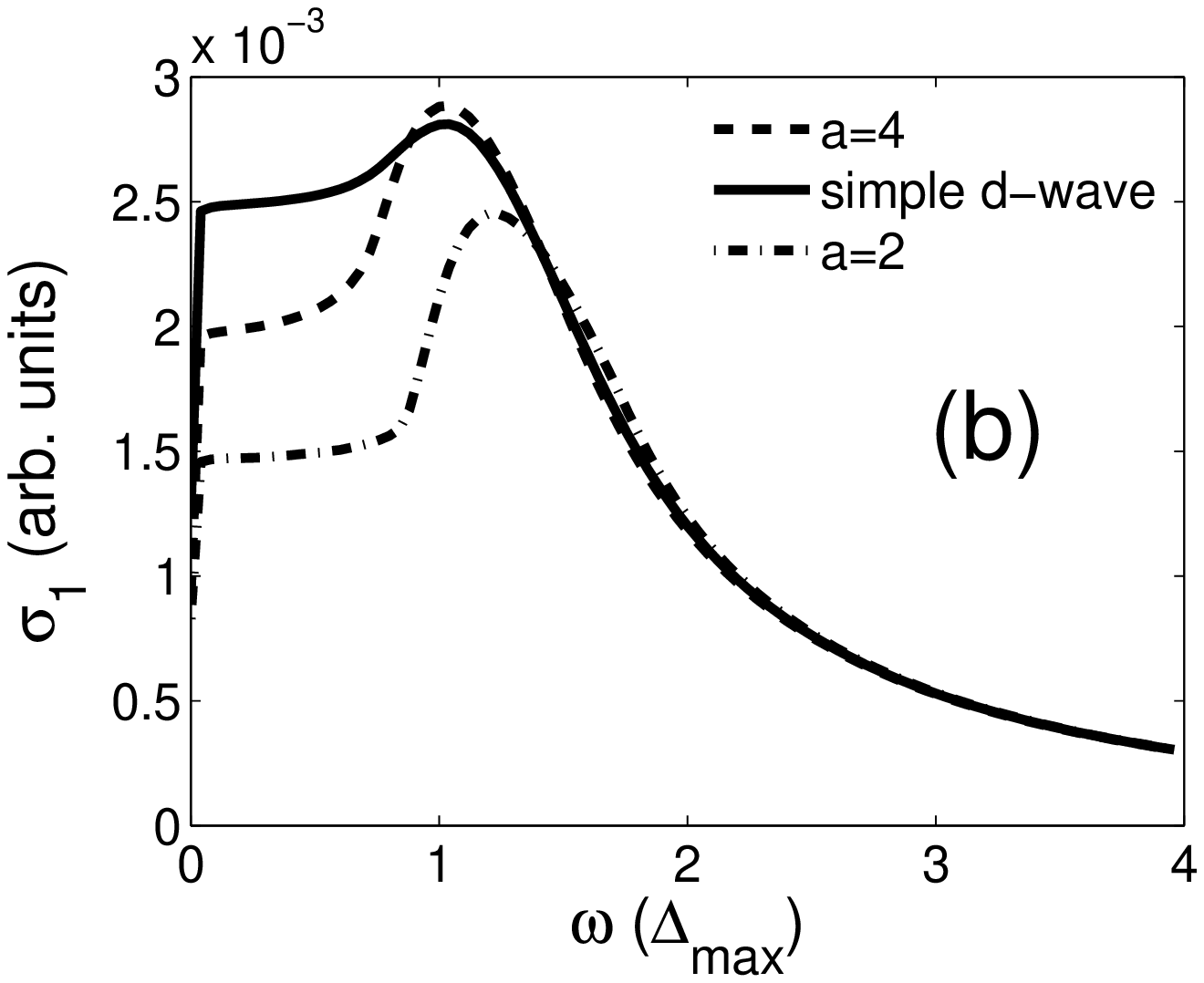,width=6cm}
\caption{ The behavior of the
optical conductivity $\sigma_1 (\omega)$ in the Born (a) and unitary
(b) limit, for a monotonic gap $\Delta (\phi) \propto \cos 2 \phi$
(solid line), and non-monotonic gap with $a=2$ (dashed-dotted line)
and $a=4$ (dashed line). } \label{fig5_1}
\end{figure}

\subsection{Optical conductivity}

Substituting the results for the self-energy into Eq (\ref{r_6}), we
obtain optical conductivity. The results are plotted in Fig.
\ref{fig5_1} (a) and (b) for Born and unitary limits,
respectively\cite{graf}. The behavior of the conductivity in the two
limits is not identical, but the interplay between the monotonic and
the non-monotonic gap is similar. In both cases, the conductivity
for a non-monotonic gap passes through a well-pronounced maximum at
some frequency below $2\Delta_{max}$,  sharply drops at smaller
frequencies, and then increases again at very low frequencies, and
at $\omega \to 0^+$ approaches the universal limit~\cite{lee} in
which the conductivity depends on the nodal velocity but does not
depend on $\gamma$ as long as  $\gamma << \Delta_{max}$. The universal
behavior is, however, confined to very low frequencies, while in a
wide frequency range below $2\Delta_{max}$ the conductivity in case
of a non-monotonic gap is strongly reduced compared to its normal
state value. The frequency at which the conductivity has a maximum
depends on $a$, and is closer to $2\Delta_{max}$ for $a=2$ than for
$a=4$.

The existence of the maximum in $\sigma_1 (\omega)$ below
$2\Delta_{max}$ can be also understood analytically. Expanding the
gap $\Delta (\phi)$ near its maximum value $\Delta_{max}$ and
substituting the expansion into (\ref{r_6}), we find, after some
algebra, that the conductivity  has a one-sided non-analyticity
below $\omega = 2\Delta_{max}$ -- it contains a negative term
proportional to $(2\Delta_{max} - \omega)^{3/2}$. This negative term
competes with a regular part of $\sigma_1 (\omega)$, which smoothly
increases with decreasing $\omega$, and gives rise to a maximum in
$\sigma_1 (\omega)$ below $2\Delta_{max}$.

The behavior of the conductivity in a superconductor with a
non-monotonic gap is  consistent with the available data on
$\sigma_1 (\omega)$ in optimally doped PCCO~\cite{homes}. The
measured conductivity has a rather strong peak at  $70 cm^{-1}$, and
decreases at smaller frequencies. The authors of Ref.
\onlinecite{homes} explained the existence of the maximum in the
optical conductivity by a conjecture  that the conductivity in a
$d-$wave superconductor with a non-monotonic gap should largely
resemble the conductivity in an $s-$wave superconductor. Our results
are in full agreement with this conjecture. The authors of Ref.
\cite{homes} also associated the peak frequency with
$2\Delta_{max}$. We found that the peak frequency is actually
located below $2\Delta_{max}$, and the difference between the two
depends on the shape of the gap. For our $a=2$, the peak frequency
is at $1.8 \Delta_{max}$ in the Born limit, and at $1.3
\Delta_{max}$ in the unitary limit. For $a=4$, the deviations are
higher.  Experimentally, $2\Delta_{max}$ in optimally doped PCCO can
be extracted from $B_{2g}$ Raman scattering (see below) and equals
$77 cm^{-1}$, see Ref. ~\onlinecite{quazil}, i.e., the peak in
$\sigma_1 (\omega)$ is at $1.8 \Delta_{max}$. This agrees with our
$a=2$ case in the Born limit.

\subsection{Raman intensity}

\begin{figure}[h]
\epsfig{file=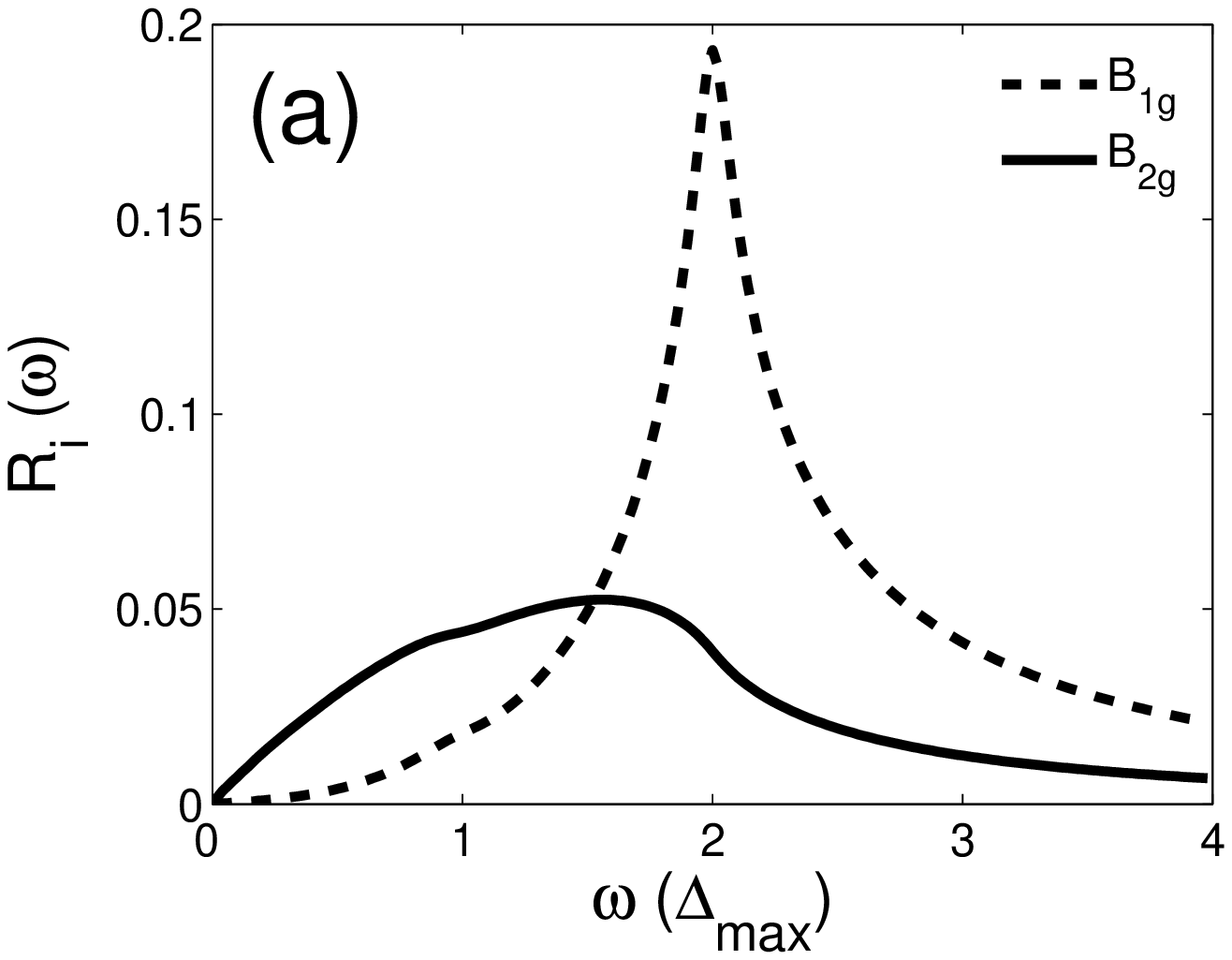,width=5cm}
\epsfig{file=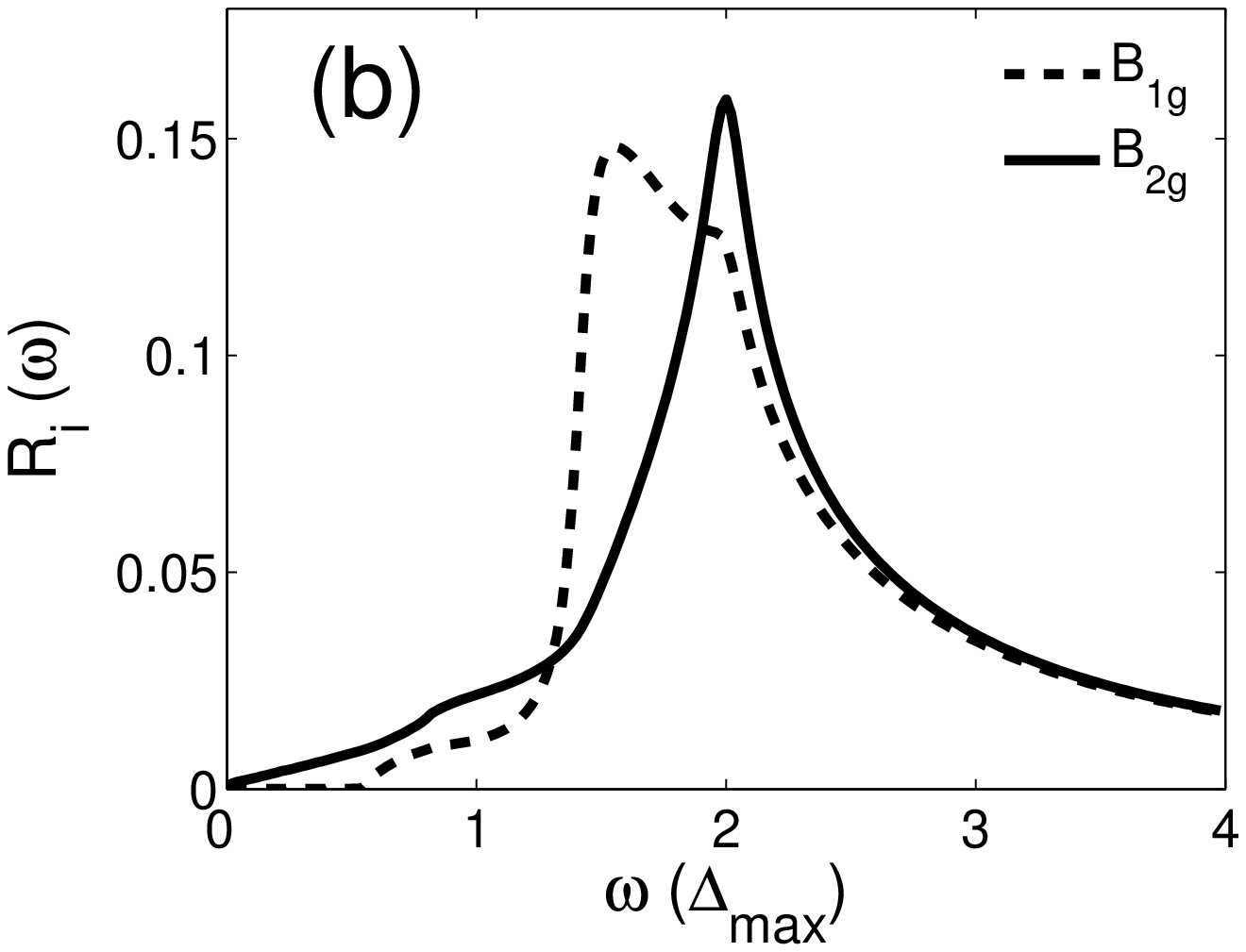,width=5cm} \hspace{-0.5cm}
\epsfig{file=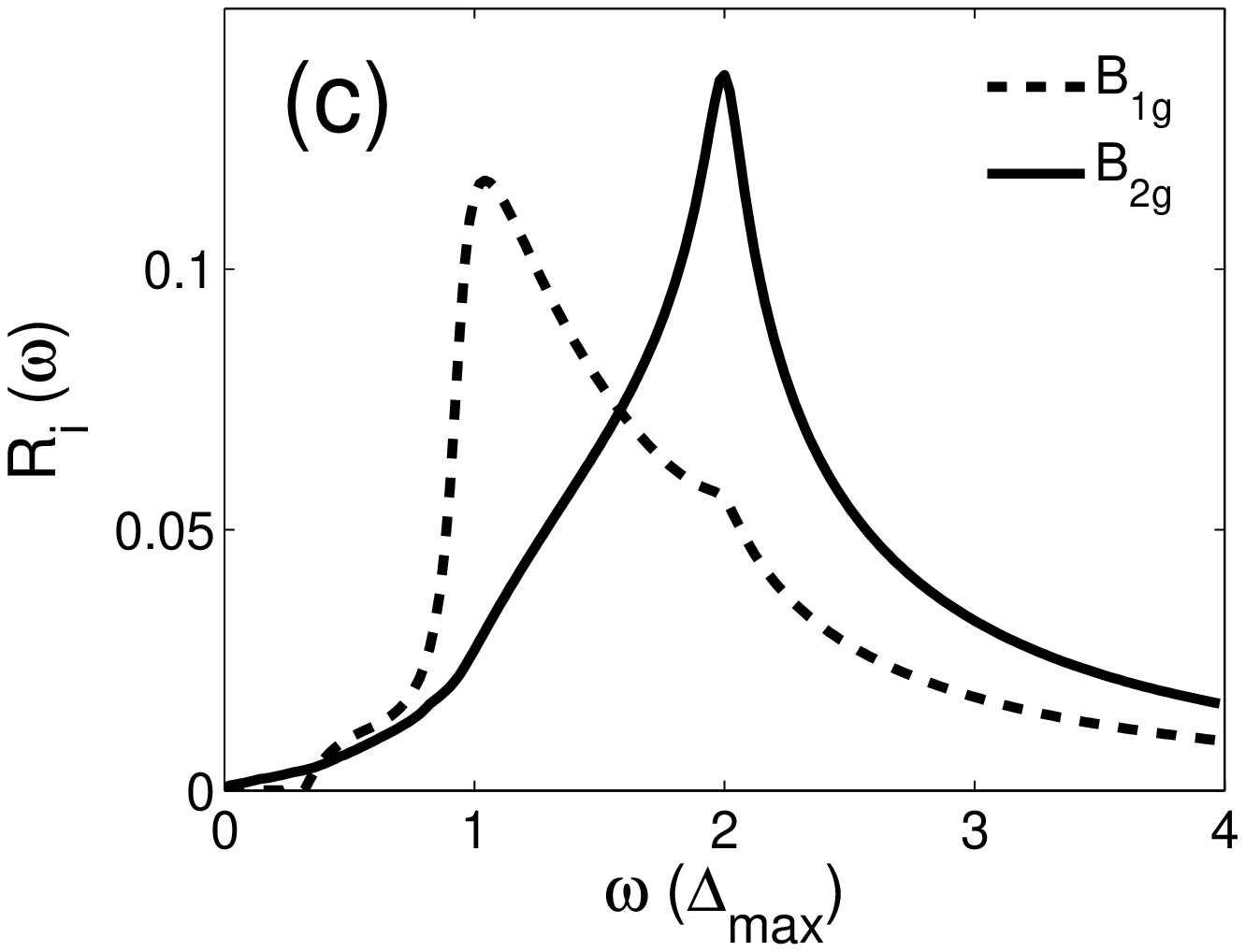,width=5cm} \caption{The behavior of the
Raman intensity $R(\omega)$ in $B_{1g}$ (dashed) and $B_{2g}$
(solid) scattering geometries for a monotonic gap $\Delta (\phi)
\propto \cos 2 \phi$ (a), non-monotonic with a=2 (b) and a=4 (c) in
the Born limit.} \label{fig5}
\end{figure}

\begin{figure}[h]
\epsfig{file=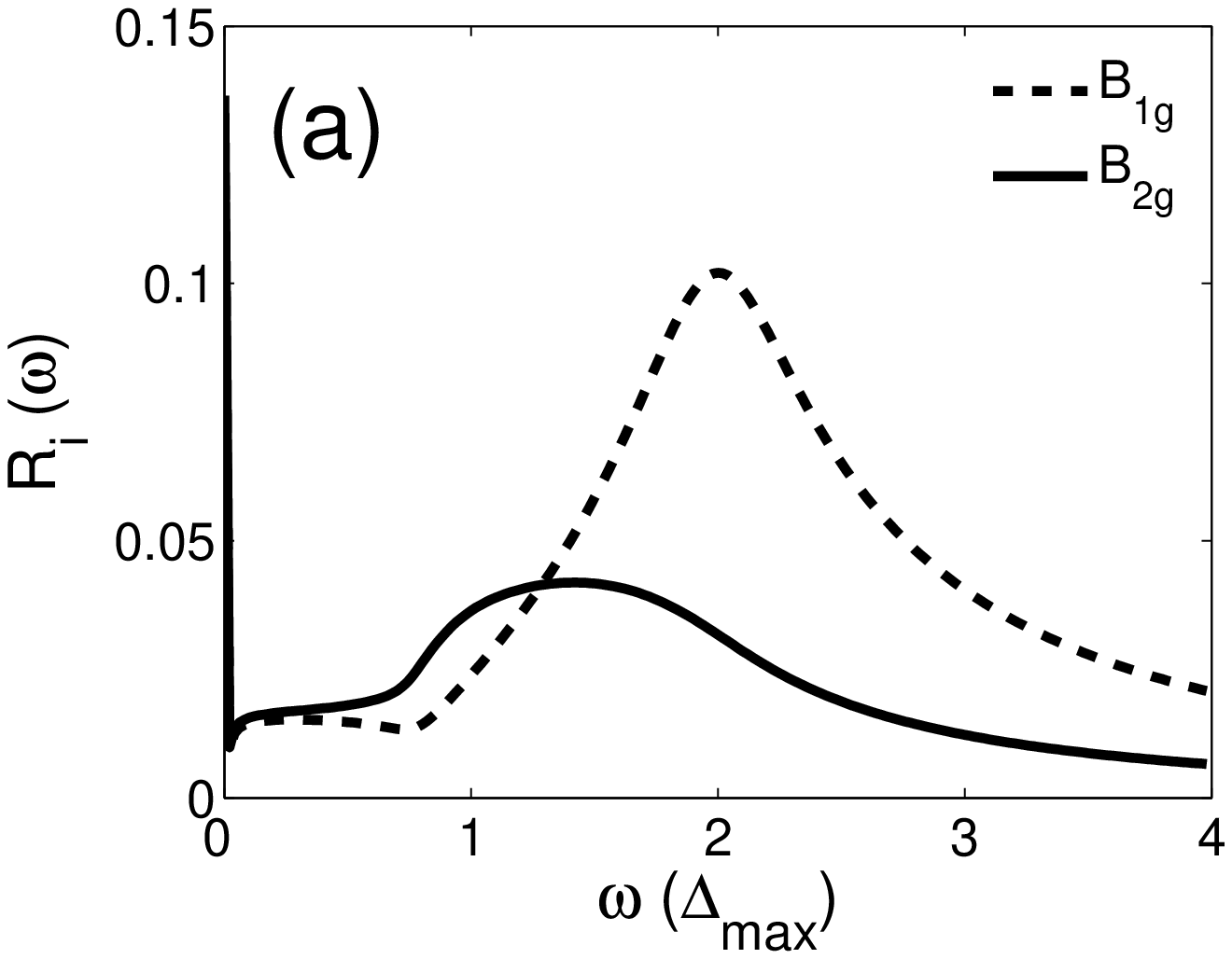,width=5cm}
\epsfig{file=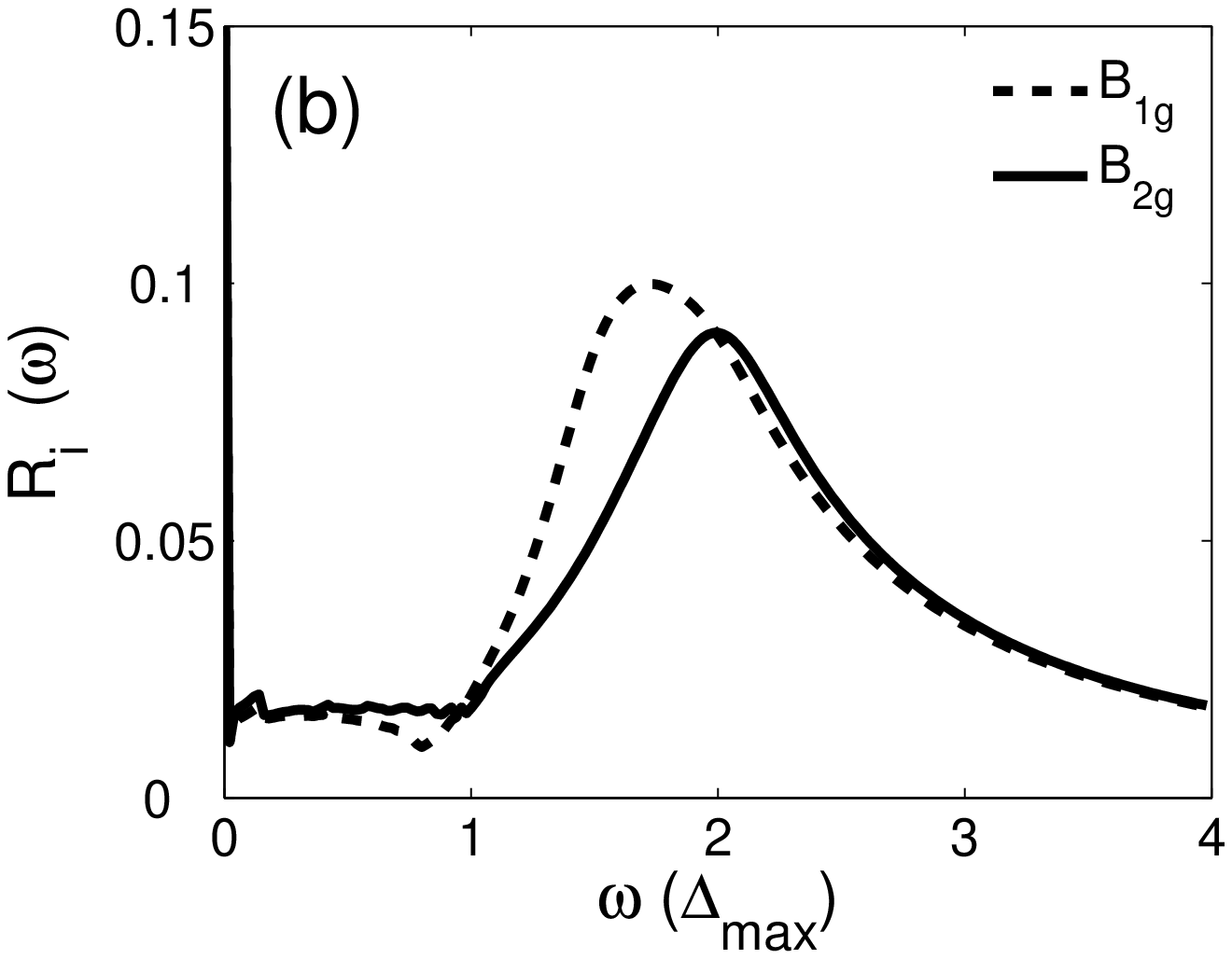,width=5cm}
\epsfig{file=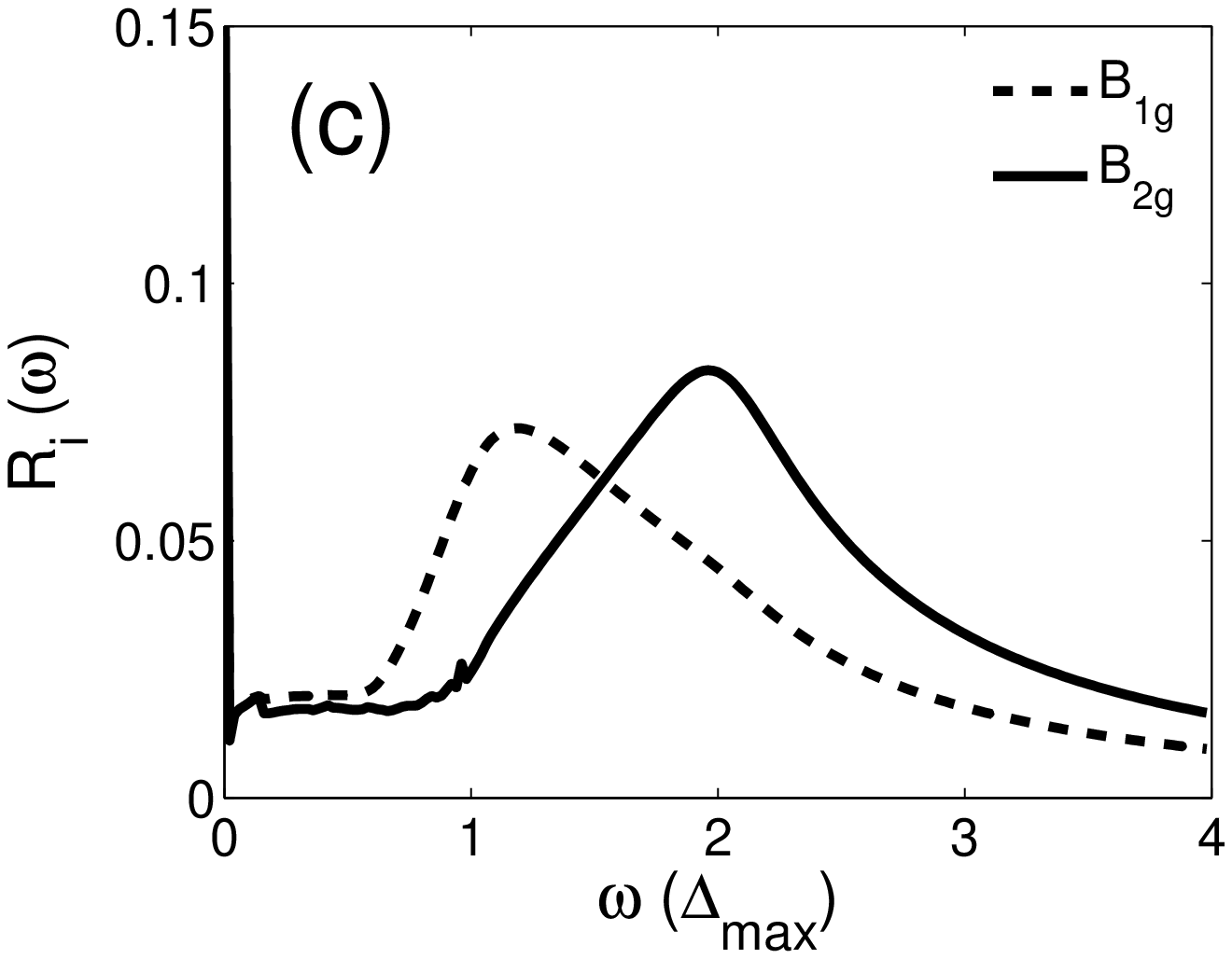,width=5cm} \caption{The behavior of the
Raman intensity $R(\omega)$ in $B_{1g}$ (dashed) and $B_{2g}$
(solid) scattering geometries for a monotonic gap $\Delta (\phi)
\propto \cos 2 \phi$ (a), non-monotonic with a=2 (b) and a=4 (c) in
the unitary limit} \label{fig6}
\end{figure}

The results for the Raman intensity are presented in Figs.
\ref{fig5}-\ref{fig6}. In a BCS superconductor with a monotonic
gap, Raman intensity has a sharp peak at $2\Delta_{max}$ in $B_{1g}$
scattering geometry, and a broad maximum at around $1.6
\Delta_{max}$ for $B_{2g}$ scattering. This behavior holds in the
presence of impurity scattering, both in Born and unitary limits,
see Figs. \ref{fig5}(a) and \ref{fig6}(a).

The $B_{1g}$ and $B_{2g}$ Raman intensities for a non-monotonic gap
are presented in Figs. \ref{fig5} -\ref{fig6} (b)-(c) for Born and
unitary limits and $a=2$ and $a=4$.  In all cases, we find the
opposite behavior: $B_{2g}$ intensity has a sharp peak at
$2\Delta_{max}$, while $B_{1g}$ intensity is very small at small
frequencies, rapidly increases around $\Delta_{max}$, passes through
a maximum,  then gradually decreases at higher frequencies and
displays a weak kink-like feature at $2\Delta_{max}$. The position
of the $B_{1g}$ peak depends on $a$ -- in both Born and unitary
limits it is close to $1.6 \Delta_{max}$ for $a=2$, and is close to
$\Delta_{max}$ for $a=4$.

The occurrence of the $'2\Delta'$ peak in $B_{2g}$ channel at a
higher frequency than in the $B_{1g}$ channel was the main
motivation in Ref.\cite{blumberg} to propose a non-monotonic
$d-$wave gap. The argument was that the gap with a maximum at
intermediate $0<\phi<\pi/4$ will have more weight in the nodal
region and less in the antinodal region, thus increasing the
effective $'2\Delta'$ for $B_{2g}$ intensity and decreasing it for
$B_{1g}$ intensity. In the optimally doped PCCO, the $B_{2g}$ peak
occurs at $77 cm^{-1}$, while the maximum  in $B_{1g}$ scattering is
around $60 cm^{-1}$. In optimally doped NCCO, the $B_{2g}$ peak
occurs at $67 cm^{-1}$, while the maximum  in $B_{1g}$ scattering is
at $50 cm^{-1}$ (Ref. \onlinecite{quazil}. The ratios of the peak
positions are $1.28$ in PCCO and $1.34$ in NCCO. This is  consistent
with our result for $a=2$ (same $a$ that gives the best fit of ARPES
and conductivity data), for which this ratio is $1.25$. Also, taking
experimental $67 cm^{-1}$ for $2\Delta$ in NCCO, we obtain
$\Delta_{max}= 4.2 meV$, in reasonable agreement with $3.7 meV$
observed in tunneling\cite{tunneling}.
In addition,  the data in Fig. 3 of Ref.\cite{blumberg}  show  that
the $B_{2g}$ peak is  nearly symmetric, while $B_{1g}$ intensity is
asymmetric around a maximum -- it  rapidly increases at frequencies
around $40 cm^{-1}$, passes through a maximum, and then gradually
decreases at higher frequencies. This behavior of $B_{1g}$ intensity
is fully consistent with Fig. \ref{fig5} -\ref{fig6} (b)-(c).

Blumberg et al.~\cite{blumberg}  also analyzed Raman intensities
at various incident photon frequencies and found resonance
enhancement of the  $B_{2g}$ intensity, but no resonance enhancement
of $B_{1g}$ intensity.  We didn't attempt to analyze the resonance
behavior of the Raman matrix element (this would require to consider
the internal composition of the Raman vertex~\cite{frenkel}). We
note, however, that the shape of the $B_{2g}$ Raman intensity
virtually does not change between the resonance and the
non-resonance cases, only the overall magnitude increases near the
resonance much like it happens in resonant Raman scattering in
insulating cuprates~\cite{bl_res}. We therefore believe that our
analysis of the Raman profile as a function of transferred frequency
is valid both in the non-resonance and in the resonance regimes.

Finally, we note that our results for $R(\omega)$  are quite similar
to Ref.\cite{rudi}, whose authors criticized the explanation of the
Raman data in terms of non-monotonic gap~\cite{comm}. However,
contrary to Ref. ~\cite{rudi}, we argue that the theoretical results
for $R(\omega)$ obtained for a non-monotonic gap agree well with the
data in Ref.\cite{blumberg}. At the same time, we agree  with Ref.
~\cite{rudi} that one can hardly extract from the data in $NCCO$ and
$PCCO$  the $\omega^3$ behavior of $B_{1g}$ intensity (which is the
Raman hallmark of $d_{x^2-y^2}$ pairing), as the low-frequency
behavior of the $B_{1g}$ intensity is dominated by a sharp increase
at frequencies of order $\Delta_{max}$.

In the analysis above we neglected final state interaction (the
renormalization of the Raman vertex). There are two reasons for
this.  For $B_{2g}$ scattering, the final state interaction is given
by the $B_{2g}$ component of the effective four-fermion interaction.
This component is repulsive, at least if the effective four-fermion
interaction comes from spin-fluctuation  exchange. The repulsive
final state interaction does not give rise to excitonic resonances,
and generally does not substantially modify the Raman
profile~\cite{convent}. For $B_{1g}$ scattering, final state
interaction is the same as the pairing interaction, {\it i.e.} it is
attractive. In general, such interaction affects the Raman
profile~\cite{cdk}. However, the interaction which gives rise to a
non-monotonic gap in the form of (\ref{r_2}) is the largest at
angles $\phi$ close enough to $\pi/4$. At these angles, the $B_{1g}$
matrix element  $\gamma_{B_{1g}} \propto \cos 2\phi$ is reduced, and
we do not expect that repeated insertions of $B_{1g}$ vertices will
substantially modify the Raman profile.

\section{conclusion}

In this paper, we  analyzed the behavior of the optical
conductivity and Raman intensity in $B_{1g}$ and $B_{2g}$ scattering
geometries in the superconducting state of electron-doped cuprates.
We  found that the results are best fitted by a non-monotonic
$d_{x^2-y^2}$ gap.  Such gap was originally suggested as a way to
explain Raman data~\cite{blumberg},  and later extracted from ARPES
measurements of the leading edge gap along the Fermi
surface~\cite{matsui}. The non-monotonic gap 
has also been obtained theoretically in the analysis of
quantum-critical pairing mediated by the exchange of overdamped spin
fluctuations~\cite{krot}.

We  found that the non-monotonic gap which  agrees best
with the ARPES data (Eq. (\ref{r_2}) with $a=2$) also fits best the
data for optical conductivity and Raman scattering. The agreement
with the data is quite good, not only in the  positions of the
maxima in optical conductivity and Raman response, but also in the
shapes of $\sigma_1 (\omega)$ and $R(\omega)$.  We argue that this
good agreement is a strong argument in favor of a non-monotonic
$d_{x^2-y^2}$ gap in electron-doped cuprates.

We thank G. Blumberg and C. Homes  for useful conversations. AVC
acknowledges support from NSF-DMR 0604406  and from
Deutscheforschungsgemeinschaft via Merkator GuestProfessorship, and
is thankful to TU-Braunshweig for the hospitality during the
completion of this work. IE is supported by the DAAD under Grant No.
D/05/50420.


\begin{thebibliography}{99}
\bibitem{comm}  The small
difference between our results and those of  Ref.\protect\cite{rudi}
are due to the fact that we used (\protect\ref{r_2}) for the gap,
while they used a $\cos 2\phi$ gap with extra higher harmonics.
\end{thebibliography}
\end{document}